\documentclass[12pt]{article}
 \pdfoutput=1
 \usepackage{graphicx}
  \usepackage{epsfig}
  \usepackage{amsmath}
  \usepackage{amssymb}
  \usepackage{color}
 \usepackage{dcolumn}
\usepackage{amsbsy}
\usepackage{bm}
\usepackage{amsthm}
\usepackage[caption=false]{subfig}
\usepackage{tabularx}
\usepackage{array}
\usepackage{hyperref}
  \newlength{\abstractwidth}
  \renewcommand{\title}[1]{\vbox{\center\bf{\Large{#1}}}\vspace{5mm}}
  \renewcommand{\author}[1]{\vbox{\center#1}\vspace{5mm}}
  \newcommand{\address}[1]{\vbox{\center\em#1}}
  \newcommand{\email}[1]{\vbox{\center\tt#1}\vspace{5mm}}

\begin{document}

\begin{titlepage}
\begin{center}
\hfill \\
\hfill \\
\vskip 1cm

\title{Statistical Mechanics Derived From Quantum Mechanics}

\author{Yu-Lei Feng$^{1}$ and Yi-Xin Chen$^1$}

\address{$^1$Zhejiang Institute of Modern Physics, Zhejiang University, Hangzhou 310027, China}

\email{11336008@zju.edu.cn} \email{yxchen@zimp.zju.edu.cn}

\end{center}

  \begin{abstract}
A pedagogical derivation of statistical mechanics from quantum mechanics is provided, by means of open quantum systems. Besides, a new definition of Boltzmann entropy for a quantum closed system is also given to count microstates in a way consistent with the superposition principle. In particular, this new Boltzmann entropy is a constant that depends only on the dimension of the system's relevant Hilbert subspace. Finally, thermodynamics for quantum systems is investigated formally.
  \end{abstract}

\end{titlepage}

\tableofcontents

\baselineskip=17.63pt








\section{Introduction}
\label{seci}

According to the number of degrees of freedom, a system is usually classified into three cases: microscopic, macroscopic and cosmoscopic. It's believed that the corresponding descriptions for those three cases are also different. For example, for a simple microscopic system quantum mechanics is enough; while for the latter two, such as a complex macroscopic system, (quantum) statistical mechanics is more useful. In (quantum) statistical mechanics~\cite{r}, Boltzmann entropy $S_{Bolt}$ is an important quantity that provides a measure of the number of microstates in a macrostate denoted by macro-quantities. Besides, thermodynamics can be obtained by means of ensemble theory, microcanonical ensemble for isolated or closed systems, (grand) canonical ensemble for open systems.

A macrostate for a macroscopic system is usually denoted as $(E,S_{Bolt},N,V)$, given by its (internal) energy $E$, Boltzmann entropy $S_{Bolt}$, particle number $N$, space volume $V$ etc. These macro-quantities can be treated as observables of the studied system, and are usually obtained through measurements. In particular in quantum statistical mechanics, quantum observables $\hat{H}$, $\hat{N}$ can be assigned so that their corresponding macro-quantities can be obtained by taking averages in terms of $Tr(\hat{\rho}_{st}\hat{O})$, with $\hat{\rho}_{st}$ the density matrix for one of those familiar ensembles\footnote{The space volume $V$ is an exception, indicating that it may be a classical quantity, just like its corresponding ``force", pressure $p$.}. Moreover, the Boltzmann entropy $S_{Bolt}$ can be expressed formally (not determined) by the von Neumann entropy $S[\hat{\rho}_{st}]$.

More specifically, the density matrix $\hat{\rho}_{st}$ in quantum statistical mechanics can be given by a general form~\cite{r}
\begin{equation}
\hat{\rho}_{st}(t)=\frac{1}{\mathcal{N}}\sum_{k=1}^{\mathcal{N}}\vert \psi^k(t)\rangle\langle \psi^k(t)\vert,
\label{5c}
\end{equation}
where the sum is over the $\mathcal{N}$ member systems of a \emph{presumed ensemble}. The state $\vert \psi^k(t)\rangle$ is some normalized quantum state for the member system $k$ at time $t$. For different ensembles, $\hat{\rho}_{st}$ will be reduced to some corresponding forms. For example, for microcanonical ensemble, according to the postulate of \emph{equal a priori probabilities}, we will have the matrix elements $(\hat{\rho}_{st})_{mn}=c\delta_{mn}$~\cite{r} in some representation. Easily to see, the ``presumed ensemble", composed of huge number of ``mental copies" of the studied system, is an independent concept. This can be seen by noting that, in the expression $Tr(\hat{\rho}_{st}\hat{O})$ there are two kinds of averages, an ensemble average and a quantum average. As a result, when calculating the von Neumann entropy $S[\hat{\rho}_{st}]$, the presupposed probability distribution $\sum_{k=1}^{\mathcal{N}}P^k=1$ for the presumed ensemble will lead to some extra \emph{fictitious information}, in addition to the quantum information encoded in the system's quantum states. The fictitious information is actually resulted from our uncertainty about the studied system. But this uncertainty is only artificial, not intrinsical for the system, especially for quantum systems. For example, a quantum closed system can be well described by a unitary evolution from some given initial state. However in statistical mechanics, microcanonical ensemble will be applied to describe a closed system. Then some fictitious information (about the uncertainty of the initial states) may be added, destroying the quantum coherence of the closed system. In a word, \emph{the Boltzmann entropy given by quantum statistical mechanics seems not to be exact, due to the implicit fictitious information of the presumed ensemble}.

Certainly, the above fictitious information is not serious for ordinary thermodynamics, in which the Boltzmann entropy is only a coarse grained quantity with additivity. This is not the case for a quantum mechanics description, in which von Neumann entropy is a fine grained quantity with only subadditivity~\cite{e,f}. In fact, thermodynamics can be treated as an effective description of the underlying quantum details. Besides, the $\hat{\rho}_{st}$ used in those familiar ensembles are actually determined according to some (extreme) conditions of the presumed ensemble. While the density matrix in von Neumann entropy must be operated always according to the quantum rules. Therefore, to obtain some more correct description, we should restrict ourselves always in the framework of quantum mechanics.

In this paper, we try to provide a derivation of statistical mechanics from quantum mechanics, so that the fictitious information from the concept of ensemble can be removed. We show that the (grand) canonical ensemble theory of statistical mechanics can be derived effectively by means of open quantum systems. In particular, the most probable distribution for statistical mechanics will correspond to some stable one for the studied open quantum system. In this way, the Boltzmann entropy for the (grand) canonical ensemble can be treated as an approximation of some entanglement entropy in the stable limit. Furthermore, a new definition of Boltzmann entropy for a quantum closed system is also given, which counts microstates in a way consistent with the \emph{superposition principle} of quantum mechanics. In particular, that new Boltzmann entropy is a constant that doesn't depend on the system's energy and space volume. In fact, it is identical to the maximum von Neumann's entropy related to the density matrixes for some measurements, thus it depends only on the dimension of the system's relevant Hilbert subspace.

This paper is organized as follows. In Sec.~\ref{secii}, some basic properties of entanglement entropy or von Neumann entropy are briefly mentioned for comparing to the Boltzmann entropy. In Sec.~\ref{seciii0}, some general investigations are given to show that the (grand) canonical ensemble theory of statistical mechanics can be derived effectively by means of open quantum systems. While in Sec.~\ref{seciii1}, a new Boltzmann entropy is defined for a quantum closed system, which is a constant independent of the system's energy and space volume. Some more details for thermodynamics of quantum systems, especially the thermal equilibrium between two macroscopic subsystems are given in Sec.~\ref{seciii3}.
In addition, two simple examples are also analyzed in Sec.~\ref{seciii2} and the Appendix~\ref{seciii4}.

\section{A Brief Introduction to Entanglement Entropy}
\label{secii}

In this intermediate section, a brief introduction to entanglement entropy is provided to compare with the Boltzmann entropy.

An entanglement entropy is defined as
\begin{equation}
S_{en}=-Tr(\hat{\rho}_{ext}\ln\hat{\rho}_{ext}),
\label{1}
\end{equation}
which is constructed from some reduced density matrix.
A reduced density matrix could occur only if the relevant system contained two or more than two independent degrees of freedom\footnote{Specially, when making a measurement on a single  quantum degree of freedom, the apparatus should also be included to form a larger closed system.}. According to the standard results of quantum mechanics, a complete collection of observables can be expressed generally as
\begin{equation}
\{\hat{O}_1,\hat{O}_2,\cdots\}, \qquad [\hat{O}_1,\hat{O}_2]=\cdots=0,
\label{3}
\end{equation}
with ``$\cdots$" denoted as the observables relevant for the rest degrees of freedom. Then the full quantum state can be expressed by
\begin{equation}
\vert\psi\rangle=\sum_{1,2,\cdots}C_{1,2,\cdots}\vert O_1,O_2,\cdots\rangle,
\label{3a}
\end{equation}
from which some reduced density matrix can be derived, for example $\hat{\rho}_{1}=Tr_{2,\cdots}\vert\psi\rangle\langle\psi\vert$.

\emph{There is not a necessary second law for entanglement entropy in the microscopic sense}. For example, consider two qubits undergoing the following two unitary processes
\begin{equation}
(\alpha\vert 0\rangle_a+\beta\vert 1\rangle_a)\vert 0\rangle_b\rightarrow\alpha\vert0\rangle_a\vert 0\rangle_b+\beta\vert1\rangle_a\vert 1\rangle_b \rightarrow (\alpha\vert 0\rangle_a+\beta\vert 1\rangle_a)\vert 0\rangle_b,
\label{4}
\end{equation}
i.e. undergoing two C-NOT gates~\cite{e,f}. We can calculate the reduced density matrix $\hat{\rho}_a$ and its corresponding entanglement entropy $S_a$. Easily to see, for the initial and final cases, $S_a=0$, while for the intermediate case, $S_a\neq0$. It is believed that the first process in Eq.~(\ref{4}) gives a second law for entanglement entropy, since the correlation is generated through the interaction~\cite{e,f}. However, for simple quantum systems the interactions which can decouple (effectively) the already correlated systems, such as the one for the second process in Eq.~(\ref{4}), also occur frequently in the microscopic sense. Therefore, no second law is necessary for entanglement entropy that can always be calculated for simple quantum systems undergoing various basic or microscopic evolutions. However, for complex macroscopic systems, the involved interactions are so complicated that \emph{the decoupling or decorrelation is usually hard in the macroscopic sense}. In this case, a Boltzmann entropy seems to be more useful, since entanglement entropy is difficult to be calculated due to the complexity. This will be discussed in the next section.

Since entanglement entropy or von Neumann entropy is wildly used in quantum information theory, its meaning is similar to the (classical) Shannon entropy. Generally speaking, it's a measure of the uncertainty before we learn a system, or a measure of how much information we have gained after we learn that system~\cite{e,f}. Roughly speaking, it provides a suitable way to quantify redundancy. This can be seen by noting the reduced density matrix in the entanglement entropy's definition, i.e. the redundancy mainly comes from the traced unobserved degrees of freedom. This meaning for the entanglement entropy seems to be different from that of the Boltzmann entropy, while the latter counts the number of microstates for a given macrostate.

Another important difference between the two kinds of entropies should be stressed: the Boltzmann entropy has additivity, while entanglement entropy or von Neumann entropy has only subadditivity. This completes the brief introduction of entanglement entropy's properties, relevant for a comparison with the Boltzmann entropy. For more detailed properties, one can consult the books~\cite{e,f}.

\section{From Quantum Mechanics to Statistical Mechanics}
\label{seciii}

As shown in Sec.~\ref{seci}, the concept of ensemble in ordinary statistical mechanics may lead to fictitious information for the studied system. To resolve this problem, in this section, we will give a pedagogical derivation of statistical mechanics according to the principles of quantum mechanics, with an emphasis on a new definition of Boltzmann entropy that is consistent with the superposition principle.

\subsection{Statistical mechanics derived from quantum mechanics}
\label{seciii0}

Since ensemble in ordinary statistical mechanics may lead to fictitious information, then \emph{whether the concept of ensemble or similar can be derived in the framework of quantum mechanics}? According to the postulate of quantum measurement~\cite{e,f}, the measurement outcomes can make up a density matrix $\hat{\rho}_{ou}$. This density matrix can be interpreted in the following way. Initially, we should prepare a sample of plenty of states to be measured. After measurement, we obtain some probability distribution on the eigenstates of the measured observable. Easily to see, the sample of states, initial or final, can be treated as some kind of ``\emph{quasi-ensemble}". The difference is that the member in the presumed ensemble is an entire system under some evolution, not some specific state.

\emph{Whether can we obtain the results of the familiar statistical ensembles by using of the above ``quasi-ensemble"}? To achieve it, we use the concept of open quantum systems or decoherence~\cite{o,p,p1}, an extension of the measurement process. Suppose the system is closed initially.
To obtain a complete description of a quantum closed system, we should find a complete set of observables as in Eq.~(\ref{3}). If the number of degrees of freedom is large enough, then the system is macroscopic according to the previous classification. In this case, there is still a complete set of observables in principle, though most of those observables are difficult to be constructed. Usually, the complete set of observables may be given by $\{\hat{H},\hat{N},\hat{O}_1,\hat{O}_2,\cdots\}$ with $\{\hat{O}_1,\hat{O}_2,\cdots\}$ denoted as the other \emph{inaccessible} observables. The system's states are given by superpositions of basis states or eigenstates for those complete set of observables, i.e.
\begin{equation}
\vert \psi(t)\rangle=\sum_{r,s,\cdots}C_{r,s,\cdots}(t)\vert E_r,N_s,\cdots\rangle,
\label{5b}
\end{equation}
where the quantum numbers for those inaccessible observables are ignored for simplicity. Then we turn on some interactions from an environment. It's assumed that the interactions are weak enough so that the the complete set of observables $\{\hat{H},\hat{N},\hat{O}_1,\hat{O}_2,\cdots\}$ can still be used approximately. Then the state of the combined system-environment is
\begin{equation}
\vert \Psi(t)\rangle=\sum_{r,s,\cdots;\chi}D_{r,s,\cdots;\chi}(t)\vert E_r,N_s,\cdots\rangle\vert \chi\rangle_E,
\label{5e}
\end{equation}
where $\vert \chi\rangle_E$ is the state of the environment, with $\chi$ denoted as the corresponding quantum numbers for short. Focusing only on the energy and particle number, after partially tracing over the environment and those inaccessible degrees of freedom, we will obtain a reduced density matrix
\begin{equation}
\hat{\rho}_{red}(t)=\sum_{r,s;\tilde{r},\tilde{s}}\rho_{r,s;\tilde{r},\tilde{s}}(t)\vert E_r,N_s\rangle\langle E_{\tilde{r}},N_{\tilde{s}}\vert.
\label{5f}
\end{equation}
For complex macroscopic systems, the involved interactions are so complicated that the off-diagonal terms in Eq.~(\ref{5f}) cannot be studied well. Fortunately, the off-diagonal terms do not contribute to the energy $E=Tr(\hat{H}\hat{\rho}_{red}(t))$ and particle number $N=Tr(\hat{N}\hat{\rho}_{red}(t))$. Under this circumstance, we can make an effective reduction
\begin{equation}
\rho_{r,s;\tilde{r},\tilde{s}}(t)\rightarrow P_{r,s}(t)\equiv\rho_{r,s;r,s}(t),
\label{5g}
\end{equation}
obtaining some probability distribution of the collection of states $\{\vert E_r,N_s\rangle\}$. This provides a ``quasi-ensemble" under some (super-operator) evolution~\cite{p1} resulted from those interactions.
Certainly, this ``quasi-ensemble" is not the presumed ensemble used in statistical mechanics. The main difference is that, in the presumed ensemble the member systems are assumed to be uncorrelated or independent~\cite{r}. While the members in the above ``quasi-ensemble" are actually correlated (weakly) through the off-diagonal terms in Eq.~(\ref{5f}).

Usually, the probability distribution $\{P_{r,s}(t)\}$ is difficult to describe, due to the complicated interactions. However, it's reasonable to propose \emph{a stability condition that the system is almost unchanged or invariant in the macroscopic sense}, provided the interactions are weak enough. In this sense, this stability condition can be expressed formally as $\delta E\simeq0,\delta N\simeq0$, with the (averaged) energy and particle number of the system given by
\begin{equation}
E=\sum_{r,s}P_{r,s}E_r, \qquad N=\sum_{r,s}P_{r,s}N_s.
\label{5h}
\end{equation}
This stability condition may also be expressed by the following relations for operators
\begin{equation}
[\hat{H},\hat{H}_{int}]\approx0,\qquad [\hat{N},\hat{H}_{int}]\approx0,
\label{5k}
\end{equation}
meaning that the interactions are so weak that their own contributions can be neglected appropriately in the macroscopic sense.
Under this stability condition, the probability distribution $\{P_{r,s}(t)\}$ will tend to some stable one $\{P^*_{r,s}\}$, up to small fluctuations or variations due to the interactions. Easily to see, the above description is analogous to the one for the grand canonical ensemble, if we set the relations $\{P_{r,s}\equiv n_{r,s}/\mathcal{N}\}$ according to probability theory. Then the enumeration of microstates is given by the weight factor~\cite{r}
\begin{equation}
W\{n_{r,s}\}=\mathcal{N}!/\prod_{r,s}(n_{r,s}!)
\label{5i}
\end{equation}
which is also the main contribution of Boltzmann entropy in statistical mechanics. The variations of $\{n_{r,s}\}$ actually come from the time dependence of $\{P_{r,s}(t)\}$ resulted from those complicated interactions. In this sense, the extreme condition of $W\{n_{r,s}\}$ used in statistical mechanics corresponds to the stability condition about the interactions in Eq.~(\ref{5k}).
And the stable distribution $\{P^*_{r,s}\}$ corresponds to the most probable distribution $\{n^*_{r,s}\}$ for the grand canonical ensemble. Then by the method of Lagrange multipliers, we can obtain the distribution for the grand canonical ensemble
\begin{equation}
\hat{\rho}_{red}(t)\stackrel{stability}{\longrightarrow}\frac{e^{-\beta(\hat{H}-\mu\hat{N})}}{Tr[e^{-\beta(\hat{H}-\mu\hat{N})}]}\equiv\hat{\rho}_{gra},
\label{5j}
\end{equation}
a reduction from the reduced density matrix $\hat{\rho}_{red}$ in Eq.~(\ref{5f}) in the stability limit. With these points, it can be concluded that \emph{the (grand) canonical ensemble can be derived effectively by means of open quantum systems}\footnote{Actually, the maximum entropy principle may be explained by the semi-group evolution of the reduced density matrix. Certainly, the derivation here is only formal, a more detailed analysis is still needed by using of the technique of open quantum systems~\cite{p1}. }.

Note that the von Neumann entropy $S[\hat{\rho}_{red}]$ for the reduced density matrix in Eq.~(\ref{5f}) is an entanglement entropy. Thus the Boltzmann entropy, expressed by $S[\hat{\rho}_{gra}]$ in terms of the grand canonical ensemble's density matrix in Eq.~(\ref{5j}), is just an approximation of the entanglement entropy \emph{near the stable point}.
Since the variations of $\{P_{r,s}(t)\}$ come from the interactions in our derivation, it thus implies that the weight factor $W\{n_{r,s}\}$ in Eq.~(\ref{5i}) is related to the interactions. In fact, the Boltzmann entropy determined by $W\{n_{r,s}\}$ is only a contribution induced by the interactions. This can be seen as follows. Note that weight factor $W\{n_{r,s}\}$ actually count the degeneracies among the distribution $\{n_{r,s}\}$, which are resulted from some weak interactions that stabilize or don't change the probability distribution $\{P_{r,s}(t)\}$. We can define a set by collecting those weak interactions satisfying the stability condition in Eq.~(\ref{5k}) as
\begin{equation}
G_{\hat{U}_{int}}=\{\hat{U}_{int}|[\hat{H},\hat{U}_{int}]\approx0, [\hat{N},\hat{U}_{int}]\approx0\}.
\label{5j'}
\end{equation}
Then we have an approximative relation
\begin{equation}
W\{n_{r,s}\}\simeq \dim G_{\hat{U}_{int}},
\label{5j''}
\end{equation}
which gives the Boltzmann entropy roughly as
\begin{equation}
S_{Bolt}\simeq k_B\ln[\dim G_{\hat{U}_{int}}].
\label{5j'''}
\end{equation}
This is a general method to define an open system's Boltzmann entropy relevant for interactions\footnote{In fact, the procedure given by Eq.~(\ref{5j'}) and Eq.~(\ref{5j'''}) can also be used for quantum closed systems, as will be shown in the next subsection.}. And a more detail expression will be given in Eq.~(\ref{26d}), with the contributions from the studied system also included.

Since the Boltzmann entropy in Eq.~(\ref{5j'''}) is relevant to the interactions with the environment, there is another one for the original closed system. In statistical mechanics, a closed system is described by a microcanonical ensemble with some fixed energy $E$, particle number $N$, etc. Moreover, its Boltzmann entropy is defined as $k_B\ln\Gamma$, where $\Gamma(E,V)=\int (d^{3N}qd^{3N}p)/h^{3}$ ($h$ is the Planck's constant) with the integral over some ``hypershell" $(E-\delta/2)\leq H(q,p)\leq(E+\delta/2)$ in the phase space~\cite{r}. Obviously, this description contains the fictitious information of the presumed ensemble, as shown below Eq.~(\ref{5c}). A quantum mechanics description is needed.

The quantum states of a quantum closed system is given by Eq.~(\ref{5b}) in terms of the eigenstates of the complete set of observables $\{\hat{H},\hat{N},\hat{O}_1,\hat{O}_2,\cdots\}$.
The system's evolution will lead to the phase change of the coefficients $\{C_{r,s,\cdots}(0)e^{-iE_rt}\}$. During this evolution, states with different $\{|C_{r,s,\cdots}(0)|\}$ will be separated and won't be correlated forever. This is different from the case of above open system, in which states with different coefficients can be correlated through the interactions with the environment. Although the choice of coefficients is uncertain, this uncertainty is not the information of the quantum closed system. Actually, that uncertainty is related to the fictitious information of the presumed ensemble in Eq.~(\ref{5c}), where different choices are incorporated into one ensemble. Thus, to avoid the fictitious information, we should always assume that the initial coefficients $\{C_{r,s,\cdots}(0)\}$ has already been chosen, though we cannot know them easily.

In fact, the information encoded in the coefficients can be acquired partly through some measurements. For each observable $\hat{O}_{i}$, there will be a corresponding density matrix for measurement outcomes $\hat{\rho}^i_{ou}$ and von Neumann entropy $S[\hat{\rho}^i_{ou}]$.
Thus we have the following measurement outcomes
\begin{equation}
\{\hat{\rho}^{\hat{H}}_{ou},\hat{\rho}^{\hat{N}}_{ou},\hat{\rho}^{\hat{O}_1}_{ou},\hat{\rho}^{\hat{O}_1}_{ou},\cdots\},\qquad \{S[\hat{\rho}^{\hat{H}}_{ou}],S[\hat{\rho}^{\hat{N}}_{ou}],S[\hat{\rho}^{\hat{O}_1}_{ou}],S[\hat{\rho}^{\hat{O}_2}_{ou}],\cdots\},
\label{5a}
\end{equation}
a complete set of density matrixes $\{\hat{\rho}^i_{ou}\}$ and their von Neumann entropies $\{\hat{S}[\hat{\rho}^i_{ou}]\}$. Theoretically, each $\hat{\rho}^i_{ou}$ can also be given by some reduced density matrix, since measurement for each observable is partial. Then their corresponding von Neumann entropies become entanglement entropies. It should be stressed that the meaning for measurement outcomes used here is always theoretical, without any measurement error. In other words, we always \emph{treat the expectation values provided by quantum mechanics as the measurement outcomes}. For example, the energy $E$ is given by the \emph{quantum average} $\langle\hat{H}\rangle$. This can also be seen from Eq.~(\ref{5h}), in which the two quantities are actually two quantum averages, with respect to the state given by Eq.~(\ref{5e}). Thus, there will be a correspondence between the complete set of observables and a complete set of macro-quantities
\begin{equation}
\{\hat{H},\hat{N},\hat{O}_1,\hat{O}_2,\cdots\}\rightleftharpoons(E,N,O_1,O_2,\cdots).
\label{5d}
\end{equation}
\emph{The meaning for ``macro" is relative to observers}, i.e. the collection $(E,N,O_1,O_2,\cdots)$ is the only macro-properties about the system that we can obtain through measurements. In this sense, we can define the microstate and macrostate in the following way. \emph{Microstates are those quantum states that contain the full quantum information of a system, while macrostates are those quantum averages, such as $(E,N,O_1,O_2,\cdots)$, which contain only probability information of the system's quantum states, with the phase information lost}\footnote{Notice that these definitions of microstate and macrostate are consistent with the descriptions of the above derived grand canonical ensemble, since the distribution $\{P_{r,s}(t)\}$ is also some probability information.}. Easily to see, microstates cannot be accessible to observers directly, while macrostates can be accessible to observers directly in the sense of measurement.
Note further that the macrostate $(E,N,O_1,O_2,\cdots)$, without ensemble average, is different from the one in ordinary (quantum) statistical mechanics. In fact, in the framework of quantum mechanics, \emph{the macrostate} $(E,N,O_1,O_2,\cdots)$ \emph{for a closed system is completely determined by some specific microstate or quantum state, in particular by the chosen initial coefficients} $\{C_{r,s,\cdots}(0)\}$.

With the above concepts of microstates and macrostates, we can define a new Boltzmann entropy for a quantum closed system, providing a measure of the number of microstates in a given macrostate $(E,N,O_1,O_2,\cdots)$. One may think that it is given by the set of von Neumann entropies $\{S[\hat{\rho}^i_{ou}]\}$ in Eq.~(\ref{5a}).
However, those von Neumann entropies contain more information (probability information) than that of only the number of microstates. Easily to see, the complete set of the maximum von Neumann entropy $\{S^i_{m}\}$ may provide the required definition. Then the total Boltzmann entropy $S_{Bolt}$ for a quantum closed system is given by $\sum_iS^i_{m}$, with the sum over the complete set of observables. In the next subsection, we will introduce a similar method, as the one given by given by Eq.~(\ref{5j'}) and Eq.~(\ref{5j'''}), to define a new Boltzmann entropy in a way consistent with the superposition principle, for any quantum closed system irrespective of the number of degrees of freedom. We also show that this new definition indeed just gives the required maximum von Neumann entropy. In fact, our defined Boltzmann entropy for a quantum closed system is determined by the number of \emph{non-vanishing} coefficients $\{C_{r,s,\cdots}(0)\}$, as shown in Sec.~\ref{seciii1}.

Usually, given a macrostate, for example $(E,N)$, states with different $\{|C_{r,s,\cdots}(0)|\}$ which \emph{correspond to different quantum closed systems} can be included\footnote{Note that the evolution of a quantum closed system only induce some phase terms, such as $e^{-iE_rt}$, without changing $\{|C_{r,s,\cdots}(0)|\}$. }. This fact can be expressed formally by the following \emph{one-to-many} correspondence between the macrostate and microstates
\begin{equation}
(E,N)\rightleftharpoons\left\{\sum_{r,s,\cdots}C_{r,s,\cdots}\vert E_r,N_s,\cdots\rangle\right\}_{E,N},
\label{5l}
\end{equation}
where the notation $\{\cdots\}_{E,N}$ means the collection of states should satisfy the constraints $E=\sum_{r,s,\cdots}|C_{r,s,\cdots}|^2E_r, N=\sum_{r,s,\cdots}|C_{r,s,\cdots}|^2N_s$. For a single quantum closed system described by some specific quantum state, this correspondence will lead to overestimate of microstates, indicating that the macrostate provides only a \emph{coarse grained} description of the quantum closed system.
Moreover, here the details for the other inaccessible observables $(\hat{O}_1,\hat{O}_2,\cdots)$ can be treated as some degeneracies. Then according to the ordinary statistical mechanics, the Boltzmann entropy is given by $k_B\ln\Gamma(E,N)$, where $\Gamma(E,N)$ is determined by counting the representative points or microstates in some relevant region of the phase space. In the present example, the relevant region is some hyperplane $(E_r=E,N_s=N)$, in which the representative points stands for the degeneracies for those inaccessible observables. Besides, the probability for each representative point is $1/\Gamma(E,N)$ according to the postulate of \emph{equal a priori probabilities} for a microcanonical ensemble. However, this kind of enumeration obviously breaks the quantum coherence of a quantum closed system, in the sense that the quantum information encoded in those coefficients $\{C_{r,s,\cdots}(0)\}$ is lost, leaving only representative points in the phase space\footnote{Note that the ordinary microstates counting is in the phase space for the microcanonical ensemble, while our counting is in the Hilbert space for the studied quantum closed system. Moreover, the phase space integral $\int (d^{3N}qd^{3N}p)/h^{3}$ is only semiclassical, since the momentum and coordinate operators are noncommutative according to quantum mechanics. }. This is because some extra postulate about the presumed microcanonical ensemble is added to ``randomize" that quantum information, so that the postulate of equal a priori probabilities is always satisfied in any representation~\cite{r}\footnote{In quantum statistical mechanics, the density matrix for a microcanonical ensemble should have the same form $(\hat{\rho}_{st})_{mn}=c\delta_{mn}$, i.e. vanishing off-diagonal terms, identical diagonal terms over the allowed range~\cite{r}, agreeing with the postulate of \emph{equal a priori probabilities}. However, for different representations, that form cannot be preserved easily. Some extra postulate is needed, namely the postulate of \emph{random a priori phases} for the probability amplitudes $\langle \phi_n\vert \psi^k(t)\rangle$, which implies that the state $\vert \psi^k(t)\rangle$ in Eq.~(\ref{5c}), for all $k$, is an \emph{incoherent} superposition of some basis $\{\vert \phi_n\rangle\}$. This extra postulate is intended to ensure noninterference among the member systems of the presumed (microcanonical) ensemble~\cite{r}. Obviously, that incoherent superposition is inconsistent with the superposition principle of quantum mechanics.}. In other words, the ordinary microstates counting is inconsistent with the \emph{superposition principle} of quantum mechanics. However, this defect is overcome in our counting, as will be shown in Sec.~\ref{seciii1}. Besides, in Sec.~\ref{seciii2} we give a concrete comparison between our defined Boltzmann entropy and the ordinary one for a single quantum particle.
Furthermore, if the closed system is open to an environment through some interactions, there will be various \emph{transitions} among those states with different $\{|C_{r,s,\cdots}|\}$ via the interactions. Among these transitions induced by the interactions, there are some which do not change the distribution $\{n_{r,s}\sim|C_{r,s,\cdots}|^2\}$, leading to the weight factor $W\{n_{r,s}\}$ in Eq.~(\ref{5i}) or the set in Eq.~(\ref{5j'})\footnote{Note that the microstates counting $W\{n_{r,s}\}$ also violates the superposition principle, in the view of breaking the correlations between the system and environment. }. This confirms the previous derivation of grand canonical ensemble around Eq.~(\ref{5j}).

In our derivation of grand canonical ensemble, the environment serves only as a particle-energy reservoir whose details are usually ignored. However, the studied system and environment are also combined to be a closed system, and can be described by quantum mechanics in principle. This can be simplified as the following example. Consider two uncorrelated closed systems whose complete quantum mechanics descriptions are given initially. Then let them interact with each other to form a new closed system. For the final closed system, its complete quantum mechanics description also exists in principle, but its complete set of observables is not known exactly, due to the complex interactions. But, if the interactions are weak enough, we can treat the original complete sets of observables for the initial two closed systems as an approximate complete set of observables for the final closed system. Different from the case of the combined system-environment, in this example both of the two (sub-)systems should be considered, without a distinction between system and environment. This situation is analyzed in Sec.~\ref{seciii3}, in particular, some kind of ``detailed balance" condition is obtained to stabilize the thermal equilibrium between two macroscopic subsystems within a larger closed quantum system.

In summary, the (grand) canonical ensemble theory of statistical mechanics can be derived effectively from quantum mechanics, by means of open quantum systems. After decohering from the environment, some kind of ``quasi-ensemble" can be obtained. However, this ``quasi-ensemble" is different from the presumed ensemble used in ordinary statistical mechanics. In particular, the presupposed probability distribution for the members in the presumed ensemble is replaced by the one reduced from the correlations between the system and environment. Thus, no fictitious information appears, since all information is about the combined system-environment. Moreover, the quantum mechanics description for a quantum closed system is also different from the one given by microcanonical ensemble. In particular, in the microcanonical ensemble description, the superposition principle is violated. This will be shown in some details in Sec.~\ref{seciii1}, with an emphasis on a new definition of Boltzmann entropy for a quantum closed system.

\subsection{Boltzmann entropy for a quantum closed system}
\label{seciii1}
According to quantum mechanics, a quantum closed system can be described in the following way. Given an initial state $\vert \phi(0)\rangle$, the system evolves in a unitary manner as $e^{-i\hat{H}t}\vert \phi(0)\rangle$, with $\hat{H}$ the total Hamiltonian of that closed system. Assuming that the Hamiltonian is the sole observable\footnote{Here, we consider only the energy for simplicity. The analysis can simply be extended to a general complete set of observables as the one in Eq.~(\ref{3}).}, the observed property of the system is given almost by its energy
\begin{equation}
E=\langle \phi(t)\vert\hat{H}\vert \phi(t)\rangle=\langle \phi(0)\vert\hat{H}\vert \phi(0)\rangle.
\label{18a}
\end{equation}
Here, \emph{energy $E$ is treated as the macro-quantity accessible to observer}, as discussed below Eq.~(\ref{5d}), even if the closed system is just a microscopic quantum system. Moreover, the macro-quantity defined as in Eq.~(\ref{18a}) is also treated as a quantum average in the sense of measurement, since we can obtain the property of a system only through some measurement\footnote{Theoretically, the quantum average in Eq.~(\ref{18a}) is well-defined in quantum mechanics. However in the practical case, when making a measurement, the studied closed system will become open to some apparatus. The studied system can be treated as quasi-closed provided that the interactions involved in the measurement satisfy the condition $[\hat{U}_{int},\hat{H}]=0$~\cite{q}, i.e. commutative with the system's Hamiltonian $\hat{H}$. This condition can be extended to the complete set of observables. In other words, the interactions should not destroy the evolution of the studied system, see Sec.~\ref{seciii3} for more details. Certainly, in the actual case that commutative condition is difficult to realize.}.
There are several entropies that can be assigned for this quantum closed system. For each instant pure state $\vert \phi(t)\rangle$, the corresponding von Neumann entropy is obviously zero. Besides, through measurements, we can also obtain the density matrix for the measurement outcomes and its corresponding von Neumann entropy. For example, if we measure the energy, then the density matrix will be
\begin{equation}
\hat{\rho}=\sum_{n}|\alpha_n|^2\vert E_n\rangle\langle E_n\vert, \qquad \vert \phi(0)\rangle=\sum_n\alpha_n\vert E_n\rangle,
\label{18b}
\end{equation}
where a discrete spectrum is assumed for simplicity. In addition, as shown in the previous subsection, we can also obtain some kind of Boltzmann entropy that corresponds to the maximum von Neumann entropy of the density matrix as in Eq.~(\ref{18b}) for some measurement outcomes. This can be seen in the following way.

Under a long enough time evolution, the states $\{\vert \phi(t)\rangle\}$ for all instants will cover a subspace of the whole Hilbert space, determined by the initial state $\vert \phi(0)\rangle$, especially the \emph{non-vanishing} coefficients $\{\alpha_n\}$ in the expansion in Eq.~(\ref{18b}). This means that \emph{for a closed system, its Hilbert space under its own evolution is only a subspace or orbit determined by some specific initial condition}. This is because those states with different $\{|\alpha_n|\}$ cannot be related by (dynamical) evolutions within the closed system. Moreover, the subspace for that system is \emph{ergodic} under further evolution.
This ergodic property for the closed system's Hilbert subspace can be well described by the following set\footnote{Condition, similar to the one in the set $G^c_{\hat{U}}$, is used in the einselection scheme as a stability criteria for interactions between the apparatus and its environment~\cite{o,p}. The method of introducing the set $G^c_{\hat{U}}$ is used in~\cite{q} to deal with quantum measurement problem.}
\begin{equation}
G^c_{\hat{U}}\equiv\{\hat{U}^c|[\hat{U}^c,\hat{H}]=0\},
\label{18}
\end{equation}
i.e. a collection of \emph{all the possible} evolutions commutative with the Hamiltonian\footnote{The condition $[\hat{U}^c,\hat{H}]=0$ is a stronger quantum version of the condition for the vanishing of energy variation $0=\delta E=\delta\langle\hat{H}\rangle$, while the latter condition is not restricted to a closed system, as shown below Eq.~(\ref{5l}). In fact, the commutative condition indicates that the evolution is just $\exp(-i\hat{H}t)$ up to some phases. Then for a closed system, the two states $\vert E\rangle$ and $c_1\vert E_1\rangle+c_2\vert E_2\rangle$ are different, in the sense that they can not be related by any (dynamical) evolution of that closed system. But for an open system, they may be related through interactions with an environment. Moreover, if $E=|c_1|^2E_1+|c_2|^2E_2$ with $E_1<E<E_2$, then the difference of the energy is zero, satisfying the condition of vanishing of energy variation.}. Here we use the superscript to emphasize that we are working in the relevant subspace of the closed system, similarly for some quantities below. Since the set $G^c_{\hat{U}}$ in Eq.~(\ref{18}) contains all the possible evolutions of the closed system, the expression $G^c_{\hat{U}}\vert \phi(0)\rangle$ just gives all the possible states of the relevant Hilbert subspace\footnote{Overall phase is also important, although it can not be observed. This means the states $e^{-i\alpha}\vert \phi\rangle$ and $\vert \phi\rangle$ can be treated as two different states since they encode different phase information.}. In this way, the relevant Hilbert subspace can be completely covered ``instantaneously", giving the maximum subspace volume. Therefore, although it may need time to form a quantum closed system in the actual case, it can still be assumed that the studied system is already closed in the theoretical analysis.

Under the evolutions in the set in Eq.~(\ref{18}), the states in the subspace $\mathcal{H}^c$ are transformed frequently as $G^c_{\hat{U}}\vert \phi(0)\rangle$, while the macro-quantity energy defined in Eq.~(\ref{18a}) is invariant. Thus we can define a quotient space as
\begin{equation}
\mathcal{H}^c_{mac}\equiv\mathcal{H}^c/G^c_{\hat{U}},\qquad \dim(\mathcal{H}^c)/\dim(\mathcal{H}^c_{mac})=\dim(G^c_{\hat{U}}),
\label{19}
\end{equation}
with the quotient space $\mathcal{H}^c_{mac}$ treated as a macro-space for macrostates denoted by macro-quantities. Easily to see, \emph{$\dim(G^c_{\hat{U}})$ just gives the number of microstates per macrostate}. Then the Boltzmann entropy can be defined as\footnote{Note that the defined Boltzmann entropy in Eq.~(\ref{20}) is similar to the one defined in Eq.~(\ref{5j'''}).}
\begin{equation}
S^c_{Bolt}=k_B\ln [\dim(G^c_{\hat{U}})]=k_B\ln [\dim(\mathcal{H}^c)],
\label{20}
\end{equation}
since $\dim(\mathcal{H}^c_{mac})=1$, that is, there is only one macrostate with some initial value $E=E_0$. Obviously,\emph{ the Boltzmann entropy defined in Eq.~(\ref{20}) for a quantum closed system is always a constant without energy and space volume dependence}, because the dimension of the relevant Hilbert subspace is fixed. Note further that the entropy defined in Eq.~(\ref{20}) is already the maximum, since the set $G^c_{\hat{U}}$ in Eq.~(\ref{18}) includes all the possible evolutions, and the relevant subspace $\mathcal{H}^c$ has already been completely covered.

Since the Boltzmann entropy defined in Eq.~(\ref{20}) is a constant, it can be treated as another macro-quantity for a quantum closed system. That is, a quantum closed system can be assigned to be at some macrostate $(E_0,S_0)$. It should be stressed that \emph{these two quantities are independent}. For instance, for two closed systems sharing the same entire Hilbert space, there will be four cases for their macrostates: (i) the same energy, different (Boltzmann) entropies; (ii) different energies, the same entropy; (iii) both energy and entropy are different; (iv) the same energy and entropy. For the last case, consider two closed systems with states $c_1\vert E_1\rangle+c_2\vert E_2\rangle$ and $c_3\vert E_3\rangle+c_4\vert E_4\rangle$, obviously they have the same (Boltzmann) entropy according to the definition in Eq.~(\ref{20}), and they may also have the same energy by suitably choosing the eigenvalues and coefficients. This means that \emph{macrostates are not suitable for denoting a quantum closed system, since most of the quantum information is hidden}. This can also be seen through the one-to-many correspondence in Eq.~(\ref{5l}) between macrostate and microstates, i.e. a macrostate can correspond to many microstates. In this sense, the approach of counting microstates for a given macrostate is not proper, because overestimate may occur when two different closed systems have the same macrostate, as shown below Eq.~(\ref{5l}).

In fact, a macro-quantity defined as a quantum average of its corresponding observable, \emph{is completely determined by the quantum states of the system, in particular by those coefficients in the superposition that encode all the quantum information}\footnote{For a complex closed system composed of lots of degrees of freedom, quantum mechanics especially the \emph{superposition principle} is still applicable, at least in principle as indicated by Eq.~(\ref{5b}).}. Moreover, the Boltzmann entropy defined in Eq.~(\ref{20}) is actually determined by \emph{the number of those non-vanishing coefficients}, i.e. the dimension of the relevant Hilbert subspace. Then one question arises. If a quantum closed system's relevant Hilbert subspace is large enough so that its dimension may even be infinite, then our defined Boltzmann entropy will also become infinite. Further, since our analysis is also suitable for simple quantum system, for example a single quantum particle, thus it indicates that the Boltzmann entropy for that particle may be infinite, violating the ordinary sense. This implies that our defined Boltzmann entropy has more meaning than that only as a measure of number of microstates. This can be explained in the following way.

Our defined Boltzmann entropy for a quantum closed system is determined by the dimension of the relevant Hilbert subspace. This can also be calculated by
\begin{equation}
S^c_{Bolt}=-k_BTr(\hat{\rho}_m\ln\hat{\rho}_m), \qquad \hat{\rho}_m=\hat{I}/d,
\label{26a}
\end{equation}
where $\hat{\rho}_m$ is a density matrix in the relevant subspace with equal probability for each eigenstate, and $d$ is the dimension of that subspace. This implies that \emph{the Boltzmann entropy defined in Eq.~(\ref{20}) is just the maximum von Neumann entropy corresponding to some measurement outcome}, confirming the argument in the last subsection. Moreover, for a general complete set of observables as in Eq.~(\ref{3}), there will be a corresponding $\hat{\rho}^{O_i}_m$ for each observable $\hat{O_i}$. Then the total density matrix is $\hat{\rho}^{O_1}_m\otimes\hat{\rho}^{O_2}_m\otimes\cdots$, with $d=d_1d_2\cdots$, leading to the addition property of Boltzmann entropy.
It should be stressed that $\hat{\rho}_m$ is usually different from the density matrix in Eq.~(\ref{18b}) which serves as some measurement outcome. Obviously, the density matrix in Eq.~(\ref{18b}) contains the probability information, while $\hat{\rho}_m$ contains little information.
In other words, the probability information encoded in the coefficients of the superposition can be acquired through von Neumann entropy. This means that, although Boltzmann entropy enumerate the microstates in a macrostate, it can not provide any meaningful quantum information about a system. To acquire quantum information, we must use von Neumann entropy for the measurement outcomes. Furthermore, as shown in Sec.~\ref{secii}, von Neumann entropy can be treated as a measure of the uncertainty before we learn a system~\cite{e,f}. Thus \emph{our defined Boltzmann entropy also provides a coarse grained measure of the uncertainty before we learn a quantum closed system}. Then, the above problem about the infinite Boltzmann entropy is resolved, in the sense that \emph{the infinite Boltzmann entropy means the largest uncertainty about a quantum closed system}. Certainly, this problem is not serious in the practical sense, since no absolutely closed system is actually present.

Here adds some notes. According to quantum mechanics, to describe a closed system completely, we should find its complete set of observables, such as the one in Eq.~(\ref{3}). Usually, the Hamiltonian is one of those observables. In the above analysis, it's assumed that the Hamiltonian is the sole observable. In general, if there are also some other (independent) observables, then degeneracy may occur. For example, an electron with spin up or down may have the same energy eigenvalue, $\{\vert E_n,\uparrow\rangle,\vert E_n,\downarrow\rangle\}$. However, this degeneracy does not means two microstates in a ``macrostate" $E_n$, since $E_n$ is just an eigenvalue, not a macro-quantity defined as a quantum average. Actually, treating an electron as a closed system, its state will be a superposition of the eigenstates $\sum_{n,s}C_{n,s}\vert E_n,s\rangle$ with $s$ denoted as the spin variable. As a result, the macrostate should be denoted as $(E,S_z)$, the (average) energy and (average) spin in the $z$ direction, since a macrostate for a quantum closed system is completely determined by its quantum state. Then to count microstates in the macrostate $(E,S_z)$, we can define a set composed of evolutions that are commutative with both the Hamiltonian and the spin observable, similar to the set in Eq.~(\ref{18}).

Generally, for a complete set of observables $\{\hat{H},\hat{O}_1,\hat{O}_2,\cdots\}$, the required set of evolutions is given by
\begin{equation}
G^c_{\hat{U}}\equiv\{\hat{U}^c|0=[\hat{U}^c,\hat{H}]=[\hat{U}^c,\hat{O}_1]=[\hat{U}^c,\hat{O}_2]=\cdots\},
\label{21b}
\end{equation}
whose dimension determines the Boltzmann entropy, according to Eq.~(\ref{20}).
In this case, the macrostate for a quantum closed system can be expressed as
\begin{equation}
(E,S_{Bolt},O_1,O_2,\cdots),
\label{21a}
\end{equation}
with $(O_1,O_2,\cdots)$ the quantum averages for the corresponding observables $\{\hat{O}_1,\hat{O}_2,\cdots\}$. Then if we measure only the energy of the system, what is the number of microstates for the energy value $E$? This often happens for macroscopic systems whose complete set of observables or macro-quantities cannot be accessible easily. According to the previous analysis, macrostates are completely determined by the system's (initial) quantum states. Thus, although the other macro-quantities are not known to us, the number of microstates is still determined by the number of non-vanishing coefficients in the initial state.

In ordinary statistical mechanics~\cite{r}, a closed system is usually described by a microcanonical ensemble, in which each ``copied" system is at some specific state in the relevant phase space, with a priori equal probability. Then according to the ergodic hypothesis, the ensemble average of \emph{any physical quantity} is identical to the long-time average of that quantity through a series of measurements on the system.
In a complete quantum mechanics description, no presumed ensemble is necessary, thus microcanonical ensemble is not suitable for a quantum closed system. However, according to the discussions around Eq.~(\ref{18}), the relevant Hilbert subspace is ergodic under the closed system's evolution. This indicates that the collection of states $\{\vert \phi(t)\rangle\}$ for all instants may serves as some ``microcanonical quasi-ensemble" for a quantum closed system. Then, according to the above average identification, for a quantum closed system, there may be a following relation for an observable $\hat{O}$
\begin{equation}
Tr(\hat{\rho}_{en}\hat{O})=\lim_{T\rightarrow\infty}\frac{1}{T}\int_{0}^{T} dt \langle \phi(t)\vert\hat{O}\vert \phi(t)\rangle.
\label{21}
\end{equation}
$\hat{\rho}_{en}$ is the density matrix for the ``microcanonical quasi-ensemble"
\begin{equation}
\hat{\rho}_{en}=\lim_{T\rightarrow\infty}\frac{1}{T}\int_{0}^{T} dt \vert \phi(t)\rangle\langle \phi(t)\vert\rightarrow\sum_{n}|\alpha_n|^2\vert E_n\rangle\langle E_n\vert,
\label{22}
\end{equation}
where $\vert \phi(0)\rangle=\sum_n\alpha_n\vert E_n\rangle$ in Eq.~(\ref{18b}) is used. Easily to see, the factor $1/T$ plays the role of equal probability, just like $1/\mathcal{N}$ in Eq.~(\ref{5c}).
In addition to the long-time average, there is also a quantum average that should be done for each instant state. For the macro-quantity energy ($E$), the above identification is applicable, in particular the reduced density matrix in Eq.~(\ref{22}) is identical to the one in Eq.~(\ref{18b}) for the measurement outcomes. However, if we measure a quantity with an observable $\hat{O}$ noncommutative with the Hamiltonian, i.e. $[\hat{O},\hat{H}]\neq0$, the evolution of the closed system will be destroyed and the reduction in Eq.~(\ref{22}) is no longer suitable. In this sense, the above identification of two averages applies only for a classical closed system, but not for a quantum closed system\footnote{This argument may be relaxed. The identification in Eq.~(\ref{21}) applies, provided that the measurements for observables noncommutative with the system's complete set of observables are forbidden.}. This confirms further that a quantum closed system can not be well described by a microcanonical ensemble, as argued in the last subsection. Meanwhile, it also implies that our defined Boltzmann entropy in Eq.~(\ref{20}) is different from the one given by ordinary statistical mechanics, as shown below Eq.~(\ref{5l}). A more concrete comparison is given in the following subsection.

\subsection{Count microstates for a single quantum free particle}
\label{seciii2}

In the last subsection, we define a new Boltzmann entropy in Eq.~(\ref{20}) for a quantum closed system, which is a constant that does not depend on the energy and space volume. This is different from the Boltzmann entropy used in ordinary statistical mechanics, which usually depends on the energy and space volume. In this subsection, we give a comparison between our Boltzmann entropy and the ordinary one in statistical mechanics for a single quantum free particle. We will show that the ordinary microstates counting given by ordinary statistical mechanics violates the superpositon principle of quantum mechanics, which is also discussed below Eq.~(\ref{5l}).

According to ordinary statistical mechanics~\cite{r}, the Boltzmann entropy for a single quantum free particle is defined as $k_B\ln\Gamma$, where $\Gamma(E)=\int (d^{3}qd^{3}p)/h^{3}$ ($h$ is the Planck's constant), and the integral is over a ``hypershell" $(E-\delta/2)\leq H(q,p)\leq(E+\delta/2)$ in the phase space. The integral over the coordinate gives the space volume $V$, while the integral over momentum gives an estimate~\cite{r}
\begin{equation}
\delta(2\pi m)^{3/2}E^{1/2}.
\label{31}
\end{equation}
Therefore, the entropy defined by $k_B\ln\Gamma$ depends on both the energy and space volume.

A quantum closed system evolves in a unitary manner. In the present example, the particle's Hamiltonian is given by $\hat{H}=\hat{\mathbf{p}}^2/2m$, its basis state is given by $\{\vert \mathbf{p}\rangle\}$. Then given an initial state, the particle will evolve as
\begin{equation}
e^{-i\hat{H}t}\sum_{\mathbf{p}}C_{\mathbf{p}}(0)\vert \mathbf{p}\rangle=\sum_{\mathbf{p}}C_{\mathbf{p}}(0)e^{-iE_{p}t}\vert \mathbf{p}\rangle.
\label{32}
\end{equation}
During a long enough time, it is believed that the phase in each coefficient has been ergodic within the interval $[0,2\pi]$. In fact, as analyzed previously, the set defined in Eq.~(\ref{18}), which contains all the possible evolutions commutative with the Hamiltonian\footnote{Certainly, those evolutions are also commutative with the momentum, since $\hat{H}=\hat{\mathbf{p}}^2/2m$ in this example. Note that the coordinate operator is not included in the complete set of observables, so our defined entropy should not depend on the space volume.}, help to cover the interval $[0,2\pi]$ ``instantaneously". Thus, the ``volume" of the Hilbert subspace for the free particle is
\begin{equation}
2\pi n,
\label{33}
\end{equation}
where $n$ is the number of the \emph{non-vanishing} coefficients $\{C_{\mathbf{p}}(0)\}$ in Eq.~(\ref{32}). Then the Boltzmann entropy is given by
\begin{equation}
k_B\ln n+k_B\ln(2\pi),
\label{34}
\end{equation}
which is independent of energy and space volume, and is consistent with the definition in Eq.~(\ref{20}) up to a universal constant.

The number $n$ may be identical to the momentum integral $\int d^{3}p$, if \emph{all the coefficients $\{C_{\mathbf{p}}(0)\}$ are non-vanishing}. This is a very special case, since the non-vanishing properties of those coefficients are uncertain. In order for $\int d^{3}p$ to be plausible, we should presume a microcanonical ensemble, in which each particle is assigned to be at some momentum state, with a priori equal probability. This presumed microcanonical ensemble may be expressed as $\hat{\rho}=\frac{1}{V(p)}\int d^{3}p\vert \mathbf{p}\rangle\langle\mathbf{p}\vert$, with $V(p)$ the volume of the momentum space. Easily to see, this microcanonical ensemble description breaks the quantum coherence in the quantum states, with the quantum information encoded in the coefficients $\{C_{\mathbf{p}}(0)\}$ randomized to become equal probability~\cite{r}. In fact, the integral $\int (d^{3}qd^{3}p)/h^{3}$ is only a semiclassical quantity for the ``copied" particle in the microcanonical ensemble, since momentum and coordinate operators are noncommutative according to quantum mechanics. Obviously, the Planck's constant $h$ is inserted because of the noncommutativity of momentum and coordinate operators. But this insertion cannot recover the full quantum description, since the \emph{superposition property} or \emph{quantum coherence} for the quantum states of the particle is absent in the integral\footnote{One may think that the particle's states can also be expressed in terms of the eigenstates of particle coordinate operator. However, in this case the superposition coefficients will be changed irregularly under the time evolution, since the particle's Hamiltonian $\hat{H}=\hat{\mathbf{p}}^2/2m$ is noncommutative with the coordinate operator. In other words, the coordinate operator is not a member of the complete set of observables.}. Therefore, the ordinary microstates counting given by ordinary statistical mechanics violates the superpositon principle of quantum mechanics, confirming the discussions below Eq.~(\ref{5l}).

Furthermore, for the quantum particle, its energy is given by the quantum average of the Hamiltonian $E=\langle\hat{H}\rangle$. Given an energy value $E_0$, one may think that all the states satisfying the condition $E_0=\sum_{\mathbf{p}}|C_{\mathbf{p}}(0)|^2E_{\mathbf{p}}$ should be included. This is not true, since overestimate may occur. This overestimate can also be expressed by an analogous one-to-many correspondence as in Eq.~(\ref{5l}), i.e.
\begin{equation}
E_0\rightleftharpoons\left\{\sum_{\mathbf{p}}C_{\mathbf{p}}(0)\vert \mathbf{p}\rangle\right\}_{E_0},
\label{35}
\end{equation}
with the notation $\{\cdots\}_{E_0}$ denoted as the collection of all the states satisfying the above energy condition. This correspondence means that there is a large number of quantum states with different $\{|C_{\mathbf{p}}(0)|\}$ that contribute to the energy $E_0$. However, states with different $\{|C_{\mathbf{p}}(0)|\}$ are always different for the (closed) quantum particle, during its own evolution $e^{-i\hat{H}t}$. In this sense, the microstates counting based only on the proposed condition $E_0=\sum_{\mathbf{p}}|C_{\mathbf{p}}(0)|^2E_{\mathbf{p}}$ is not suitable. Moreover, when making a measurement on a quantum closed system, to avoid overestimate, the interactions between the system and the apparatus should satisfy $[\hat{U}_{int},\hat{H}]=0$, an extension of the condition in the set defined in Eq.~(\ref{18}). Certainly, that condition is meaningful only in a theoretical sense, since there is not an absolutely closed system in an actual case.

In general, to obtain a correct counting of microstates for a quantum closed system, the superposition principle should be preserved so that the number of microstates is given by the number of non-vanishing superposition coefficients. Besides, in the present example,
all of those states satisfying the energy condition $E_0=\sum_{\mathbf{p}}|C_{\mathbf{p}}(0)|^2E_{\mathbf{p}}$ might be included, only if they were related by some (dynamical) evolutions that can only be induced by interactions with an extra environment, as shown below Eq.~(\ref{5l}).
Further in nature, a system is always open to various environments, then its initial information is usually already ``lost" due to complex interactions. This \emph{absolute uncertainty or randomization} may lead to the presumed microcanonical ensemble, together with the postulate of equal a priori probabilities. In this case, the phase space integral $\int (d^{3}qd^{3}p)/h^{3}$ is applicable, but the energy condition may be perturbed by the (weak) interactions with those environments, with some small uncertainty $\delta$. Hence, the definition $k_B\ln\Gamma$ with integral over a hypershell may be treated as the Boltzmann entropy for a canonical ensemble. The canonical ensemble can be effectively described by the concepts of open quantum systems, as will be shown in the next subsection.

\subsection{Thermodynamics and statistical mechanics for quantum systems}
\label{seciii3}
According to the discussions in Sec.~\ref{seciii0}, statistical mechanics can be derived by means of open quantum systems. Then, thermodynamics can be derived from statistical mechanics, with temperature emerged as some multiplier~\cite{r}. For a quantum closed system with a complete quantum mechanics description, whether thermodynamics could also be derived? If the answer was yes, then we could define thermodynamics for any quantum system, even for a single quantum particle whose Boltzmann entropy is well given by Eq.~(\ref{34}). In this subsection, we still only consider the energy for simplicity, and try to find the concept of temperature for a general quantum system.

Thermodynamics is a theory for those macro-quantities or macrostates. According to Sec.~\ref{seciii1}, for a quantum closed system, its macrostate is given by two independent quantities $(E,S_{Bolt})$. Obviously, given only this macrostate, no thermodynamics can be derived since no temperature can be defined in this case. However, macrostates can change if \emph{interactions} with some other system are turned on. Assuming two initially independent closed systems $a$ and $b$, the total Boltzmann entropy for these two uncorrelated systems is
\begin{equation}
S^c_a+S^c_b=k_B\ln[\dim(\mathcal{H}^c_a)\dim(\mathcal{H}^c_b)],
\label{23}
\end{equation}
according to Eq.~(\ref{20}). Then let $a$ and $b$ interact with each other. Under those interactions, the two systems are combined to become a new closed system with a corresponding Boltzmann entropy
\begin{equation}
S^c_{ab}=k_B\ln[\dim(\mathcal{H}^c_{ab})].
\label{24}
\end{equation}
Then the change of the Boltzmann entropy due to the interactions is given by
\begin{equation}
\Delta S_{int}=S^c_{ab}-(S^c_a+S^c_b)=k_B\ln[\dim(\mathcal{H}^c_{ab})/\dim(\mathcal{H}^c_a)\dim(\mathcal{H}^c_b)],
\label{25}
\end{equation}
depending on the dimensions of the initial and final Hilbert subspaces, $\mathcal{H}^c_a\otimes\mathcal{H}^c_b$ and $\mathcal{H}^c_{ab}$. Obviously, the dimension of $\mathcal{H}^c_{ab}$ depends on the details of the interactions included in the combined system's full Hamiltonian, i.e. $\hat{H}_{ab}=\hat{H}_{a}+\hat{H}_{b}+\hat{H}_{int}$. During the process, there is also a corresponding energy change
\begin{equation}
\Delta E_{int}=E^c_{ab}-(E^c_a+E^c_b),
\label{25a}
\end{equation}
with $E^c_{a}$, $E^c_{b}$ and $E^c_{ab}$ the (average) energies for the initial closed systems $a$, $b$ and the final combined closed system. Given these changes for those macro-quantities, the microscopic details of those interactions can be effectively described by a first law
\begin{equation}
\Delta E_{int}=T\Delta S_{int},
\label{25b}
\end{equation}
with temperature $T$ defined as a proportional parameter. Therefore, for a quantum closed system alone, no thermodynamics can be derived. However, if macrostates change due to some interactions, thermodynamics can be derived in terms of the first law in Eq.~(\ref{25b}), which is also the definition of temperature $T$. Obviously, the interactions are crucial for thermodynamics, since they lead to the \emph{meaningful changes} for macrostates\footnote{If without interactions, macro-quantities are added trivially. Then for system $a$, the changes of its macrostate is just the macrostate of system $b$, which should not be used to defined temperature, otherwise thermodynamics could be derived for the quantum closed system $b$.}.

The above analysis is actually \emph{universal} for both simple microscopic and complex macroscopic quantum systems in the ordinary sense. Since the details of interactions are not known, we can only give some qualitative discussions. Assuming that the temperature $T$ in Eq.~(\ref{25b}) has the meaning of thermal temperature, i.e. $T\geq 0$, then there will roughly be two cases: (i) $\Delta E_{int}\geq0,\Delta S_{int}\geq0$, some possible repulsive interactions; (ii) $\Delta E_{int}<0,\Delta S_{int}<0$, some possible attractive interactions. In this sense, the case (i) may correspond to particle collisions that produce ordinary heat, while the case (ii) seems to be useful in forming matter's structures. Certainly, if $T$ is not a thermal temperature, then it may be negative $T<0$, and more cases will occur. These still need more investigations. In the subsequent discussions, we assume $T$ is a thermal temperature.

Whether we can distinguish $a$ from $b$ within the combined closed system, if $a$ and $b$ are distinguishable initially? The property of the formed new closed system depends on the interactions. If the (attractive) interactions are strong enough so that the two systems could produce collective modes, then $a$ and $b$ will be indistinguishable, and the combined closed system can only be described by the total Hamiltonian $\hat{H}_{ab}$. If the interactions' strengths are different, it's reasonable that there would be a fuzzy boundary between the two systems' Hilbert subspaces\footnote{The ``boundary" is actually a region in the Hilbert space, but it is small when comparing with the two systems' ``bulk" regions.}. Far from that boundary, the interactions are weak so that the two systems seems to evolve roughly independently. While near that boundary, the interactions are strong enough that collective modes could be produced.

The case for strong interactions is usually difficult to be described, so we assume that the interactions are weak enough. If system $a$ is the studied system, then system $b$ may be treated as an environment, i.e. system $a$ become an open system. There may be two situations. The first one involves the condition $[\hat{H}_{a},\hat{H}_{int}]=0$, under which the evolution of system $a$ is preserved, and its Boltzmann entropy is the same as the one $S^c_a$ for the initial closed system. This commutative condition applies well to the case of quantum measurement~\cite{o,p,q}, where the probability information encoded in the coefficients of the initial state is unchanged. The second situation is the one in which the involved interactions are only weak perturbations. Under this condition, the initial states for system $a$ may be transformed into other states, or the original spectrum for system $a$ may be split into some sub-spectrum, etc. As a result, the relevant Hilbert subspace for system $a$ will be enlarged a little, and the corresponding Boltzmann entropy increases\footnote{As shown previously, if the interactions are attractive, the Boltzmann entropy may decrease. Here, we assume that the attractive interactions have been ``exhausted" in forming matter's structures, and the rest interactions come mainly from the collisions that produce heat. }. The familiar example for this situation is the (grand) canonical ensemble in statistical mechanics, after partially tracing over the environment $b$ whose role is just a particle-energy reservoir, as already shown in Sec.~\ref{seciii0}.

Now, let's consider the open macroscopic system $a$ interacting weakly with the macroscopic environment $b$ in some quantitative way. Since the involved interactions are weak, $\{\hat{H}_a,\hat{H}_b\}$ can serve as an approximate complete set of observables of the combined closed system. Thus we can define some macro-quantities approximately as
\begin{equation}
E_a=~_{ab}\langle\hat{H}_a\rangle_{ab}, \qquad E_b=~_{ab}\langle\hat{H}_b\rangle_{ab}, \qquad E_a+ E_b\simeq E^c_{ab},
\label{25c}
\end{equation}
where the averages are over the quantum states of the combined $ab$ closed system. The last relation indicates that the interaction energy $_{ab}\langle\hat{H}_{int}\rangle_{ab}$ is small compared with the energies of the (macroscopic) systems $a$ and $b$. There are also Boltzmann entropies for the systems $a$ and $b$
\begin{equation}
S_a, \qquad S_b, \qquad S_a+ S_b\simeq S^c_{ab},
\label{25d}
\end{equation}
whose expressions are given by Eq.~(\ref{26d}) below. Notice that the relation $S_a+ S_b\simeq S^c_{ab}$ for the Boltzmann entropies is different from the case of general von Neumann entropies, since according to Eq.~(\ref{20}) $S^c_{ab}$ depends only on the dimension of the relevant Hilbert subspace\footnote{Note that, since $\{\hat{H}_a,\hat{H}_b\}$ serves as an approximate complete set of observables, then a general state of the combined system can be expressed as $\sum_{n,i}C_{n,i}\vert n\rangle_a\vert i\rangle_b$.}. By using of Eqs.~(\ref{25})~(\ref{25a})~(\ref{25c})~(\ref{25d}), we will have
\begin{equation}
\Delta E_a+\Delta E_b\simeq\Delta E_{int}, \qquad \Delta S_a+ \Delta S_b\simeq \Delta S_{int},
\label{25e}
\end{equation}
where $\Delta E_a=E_a-E^c_a$ and $\Delta S_a=S_a-S^c_a$, similarly for the quantities of system $b$. We can propose the first law for systems $a$ and $b$ respectively as
\begin{equation}
\Delta E_a=T_a\Delta S_a, \qquad \Delta E_b=T_b\Delta S_b,
\label{25f}
\end{equation}
with two temperatures $T_a$ and $T_b$. Then by using of Eqs.~(\ref{25b})~(\ref{25e})~(\ref{25f}), we have
\begin{equation}
(T-T_a)\Delta S_a+(T-T_b)\Delta S_b\simeq0,
\label{25g}
\end{equation}
which leads to $T\simeq T_a\simeq T_b$ since $\Delta S_a$ and $\Delta S_a$ are independent\footnote{Note that the relation $\Delta S_a+ \Delta S_b\simeq \Delta S_{int}$ in Eq.~(\ref{25e}) is not a constraint, but an approximate function $\Delta S_{int}(\Delta S_a,\Delta S_b)$ of two independent variables.}. This implies that the combined closed system $ab$ is homogeneous or (thermal) equilibrium between $a$ and $b$ in the macroscopic sense.

After the equilibrium between $a$ and $b$ is achieved, there are still variations for those macro-quantities due to the weak interactions. For example, for system $a$ we have
\begin{equation}
\delta E_a=\delta~_{ab}\langle\hat{H}_a\rangle_{ab}=~_{ab}\langle\delta\hat{H}_a\rangle_{ab}\simeq~_{ab}\langle[\hat{H}_{a},\hat{H}_{int}]\rangle_{ab},
\label{26b}
\end{equation}
similarly for the Boltzmann entropies $S_a$ due to the variation of the density matrix for measurement outcomes. Since $E_a=E^c_a+\Delta E_a$ and $E^c_a$ is a constant for quantum closed system, we have $\delta E_a=\delta (\Delta E_a)$, similarly for other quantities. Then from Eq.~(\ref{25e}), we will have\footnote{Certainly, Eq.~(\ref{26f}) can also be derived directly from the variations of the relations in Eqs.~(\ref{25c}) and~(\ref{25d}), meaning the \emph{measured} total energy and entropy is almost invariant near the stable point. }
\begin{equation}
\delta E_a+\delta E_b\simeq0, \qquad \delta S_a+ \delta S_b\simeq 0,
\label{26f}
\end{equation}
where $\delta(\Delta E_{int})=\delta(\Delta S_{int})=0$ has been used because they are constants, as indicated by Eqs.~(\ref{25}) and~(\ref{25a}). The two relations in Eq.~(\ref{26f}) can be regarded as some kind of ``\emph{detailed balance}" conditions between $a$ and $b$ in terms of variations of macro-quantities. Analogously, by making variations of Eqs.~(\ref{25f}) and~(\ref{25g}), we obtain that the first law and the relation $T\simeq T_a\simeq T_b$ is stable, implying that \emph{the (thermal) equilibrium between systems $a$ and $b$ is stable}.

In most cases, the environment $b$ is usually not important, and we can focus on the studied system $a$ alone. Then, for a stable (thermodynamic) equilibrium, we can propose some stability condition $[\hat{H}_{a},\hat{H}_{int}]\approx0$ as in Eq.~(\ref{5k}), so that the variations of the macro-quantities for system $a$ would vanish approximately, for example $\delta E_a\simeq0$. This means that the macrostates for the (macroscopic) system $a$ are almost invariant.
Analogous to the set $G^c_{\hat{U}}$ defined in Eq.~(\ref{18}) for a quantum closed system, we can also define a set of interactions as
\begin{equation}
G_{\hat{U}_{int}}\equiv\{\hat{U}_{int}|[\hat{U}_{int},\hat{H}_a]\approx0\}.
\label{26c}
\end{equation}
i.e. a collection of \emph{all the possible} interactions $\hat{U}_{int}\simeq e^{-i\hat{H}_{int}t}$ that are (negligible) perturbations of system $a$. Since the macro-quantities are almost invariant under those interactions in Eq.~(\ref{26c}), we can also define an approximate quotient space like the one in Eq.~(\ref{19}) and obtain
\begin{equation}
S_a\approx k_B\ln[\dim(\hat{P}_aG_{\hat{U}^c_{a}}G_{\hat{U}_{int}})],\qquad S^c_a=k_B\ln[\dim(G_{\hat{U}^c_{a}})],
\label{26d}
\end{equation}
where a projector $\hat{P}_a$ is used to subtract the contribution from the environment $b$ due to those interactions. Here, $G_{\hat{U}^c_{a}}\equiv\{\hat{U}^c_{a}|[\hat{U}^c_{a},\hat{H}_a]=0\}$ is the collection of all the possible evolutions of the original closed system $a$, and the elements in $G_{\hat{U}^c_{a}}G_{\hat{U}_{int}}$ are of the form $\hat{U}^c_{a}\hat{U}_{int}\simeq e^{-i(\hat{H}_a+\hat{H}_{int})t}$. In this way we obtain the entropy change due to the interactions
\begin{equation}
\Delta S_a=S_a-S^c_a\approx k_B\ln[\dim(\hat{P}_aG_{\hat{U}_{int}})],
\label{26e}
\end{equation}
similarly for the case of the environment $b$ with a corresponding projector $\hat{P}_b$\footnote{From Eq.~(\ref{25e}), we have the full entropy change $\Delta S_{int}\approx k_B\ln[\dim(G_{\hat{U}_{int}})]$, where $\hat{P}_a+\hat{P}_b\simeq\hat{I}$ has been used. Obviously, the two projectors breaks the correlations between systems $a$ and $b$, confirming that the weight factor $W\{n_{r,s}\}$ in Eq.~(\ref{5i}) also violates the superposition principle.}. Analogous to the stability condition $\delta E_a\simeq0$, there is also a corresponding one for the Boltzmann entropy $S_a$. The two stability conditions can be expressed as
\begin{equation}
\delta S_a=\delta(\Delta S_a)\simeq 0, \qquad \delta E_a=\delta(\Delta E_a)\simeq0,
\label{27a}
\end{equation}
where the relations $S_a=S^c_a+\Delta S_a,E_a=E^c_a+\Delta E_a$ and $\delta S^c_a=0,\delta E^c_a=0$ have been used. With the help of the method of \emph{Lagrange multipliers}, we have $\delta(\Delta S_a-T_a^{-1}\Delta E_a)\simeq0$ with the thermal temperature emerging as a multiplier, which is consistent with the first law in Eq.~(\ref{25f}). This is analogous to the familiar canonical ensemble of statistical mechanics, i.e. finding the most probable or the stable distribution that maximizes the Boltzmann entropy, together with the energy constraint. This analysis confirms the discussions in Sec.~\ref{seciii0}, where the grand canonical ensemble is also derived by means of open quantum systems. Moreover, the Boltzmann entropy change $\Delta S_a$ given by Eq.~(\ref{26e}) is related to the interactions evidently, just like the one in Eq.~(\ref{5j'''}. This indicates further that the weight factor $W\{n_{r,s}\}$ in Eq.~(\ref{5i}) or its corresponding Boltzmann entropy is indeed a contribution induced by the interactions.

By comparing Eqs.~(\ref{26f}) with~(\ref{27a}), we can see that the latter stability condition for system $a$ alone is just a special case of the former one, if analogous conditions as in Eq.~(\ref{27a}) are also proposed for the environment $b$, i.e. $\delta E_b\simeq0,\delta S_b\simeq0$. In ordinary statistical mechanics, the environment is usually ignored so that the condition in Eq.~(\ref{27a}) can be proposed approximately. This was possible only if the stability condition for the environment were presupposed. However, if we consider two macroscopic subsystems within a larger closed system, none of them should be ignored. Under this circumstance, the ``detailed balance" conditions between those two subsystems, such as the one in Eq.~(\ref{26f}), should also be added. This may also be explained in the following way. The stability condition for each subsystem only gives its own temperature, such as $\delta(\Delta S_a-T_a^{-1}\Delta E_a)\simeq0$, and $\delta(\Delta S_b-T_b^{-1}\Delta E_b)\simeq0$. Without the ``detailed balance" conditions, the two temperatures cannot be identified, and the thermal equilibrium cannot be established effectively. Furthermore, for a more larger closed system composed of more than two (macroscopic) components, the ``detailed balance" conditions between any two components may be proposed approximately, provided the stability conditions for the rest components could be presupposed appropriately. Certainly, this depends on the details of the relevant interactions, and needs further investigations.

Here adds some notes. Although the above macro-quantities are usually defined as quantum averages like the one in Eq.~(\ref{18a}), their relations are also applicable classically. For example, the first law has the similar form as that of ordinary thermodynamics. In this sense, we can conclude that thermodynamics and statistical mechanics indeed can be derived from quantum mechanics step by step. Further, the thermal meaning of the temperatures $T$, $T_a$ and $T_b$ can also be seen in the following way. The relations in Eq.~(\ref{25e}) are unchanged by adding extra terms $0=\Delta Q_a+\Delta Q_b$ and $0=\Delta S^Q_a+\Delta S^Q_b$, which are contributions of the classical heat. This means that there are classical heat and entropy flows between $a$ and $b$.
Under this circumstance, the thermal temperature between them is given by $T_a=\Delta Q_a/\Delta S^Q_a=\Delta Q_b/\Delta S^Q_b=T_b$. This can also be treated as the definition of thermal temperature in a measurement sense, if system $b$ is some thermometer whose perturbations on the studied system can be neglected. These confirm that $T$ defined in Eq.~(\ref{25b}) indeed can be treated as a thermal temperature. The distinction is that $T_a$ or $T_b$ has more practical meaning, because only local measurements as in Eq.~(\ref{25c}) have been performed. But $T$ also applies to a general process even the involved interactions are strong, while $T_a$ or $T_b$ is meaningful only if the interactions are weak enough so that the change $\Delta E_a$($\Delta S_a$) can even be replaced by a differential $dE_a$($dS_a$), leading to the first law $dE_a=T_a dS_a$ in the differential form.

The above derived thermodynamic rules are almost formally similar to the ordinary one. In particular, the ``detailed balance" conditions in Eq.~(\ref{26f}) between systems $a$ and $b$ should also be proper for ordinary thermodynamics, with the Boltzmann entropies given by the approximation of the corresponding entanglement entropies in the stable limit. This is because the ``detailed balance" conditions are also derived in the stable limit, according to some stability requirement that the observed macro-quantities for a closed system should be almost invariant in its (internal) thermal equilibrium. The physical meanings for those ``detailed balance" conditions are easy to understand. The first energy condition says that the total energy of the entire closed system is (almost) stable, while the latter one for the Boltzmann entropy just gives the maximum entropy condition. In general, \emph{to achieve a stable thermal equilibrium, there should be three (stability) conditions: the first law as in Eq.~(\ref{25f}), the ``detailed balance" conditions as in Eq.~(\ref{26f}) and the equal thermal temperature}\footnote{Note that the ``detailed balance" condition for the Boltzmann entropy can be derived according to the other conditions. This confirms the maximum entropy condition for the whole closed system.}. Certainly, these conditions apply only in the stable limit, giving an approximative description. The more exact and complicated descriptions can only be well given by quantum rules\footnote{In a full quantum mechanics description, the complete set of density matrixes and von Neumann entropies in Eq.~(\ref{5a}) should be used.}. In ordinary thermodynamics or statistical mechanics, only the first and the last conditions are proposed, with the second ``detailed balance" condition negelected\footnote{In ordinary statistical mechanics, the ``detailed balance" conditions cannot be simply derived since the environment is usually ignored. In addition, the ``detailed balance" condition for the entropy is more strict than the familiar second law $\delta S\geq0$ for a (quantum) closed system, since we are always working in the stable limit region $\delta S\simeq0$.}.

\section{Conclusions}
\label{secv}

From the above analysis, it can be concluded that statistical mechanics can be derived effectively from quantum mechanics by means of open quantum systems. As a consequence, the fictitious information resulted from the concept of ensemble can be removed, so that the Boltzmann entropy gives only the intrinsical information of the studied open system together with some possible environment. In other words, the uncertainty of the studied open system is completely from the interactions with some environment.

Besides, a new definition of Boltzmann entropy for a quantum closed system can be given to count microstates in a way consistent with quantum coherence or superposition property of quantum states. In particular, this new Boltzmann entropy is a constant that depends only on the dimension of the system's relevant Hilbert subspace. In other words, a quantum closed system cannot be described well by microcanonical ensemble theory which may lead to fictitious information.

\appendix

\section{Statistical mechanics for an ideal quantum gas}
\label{seciii4}

In Secs.~\ref{seciii0} and~\ref{seciii3}, the (grand) canonical ensemble theory of statistical mechanics is derived effectively by means of open quantum systems. In this appendix, a
simplified open quantum system is studied to derive the familiar distribution for the indistinguishable quantum particles.

Consider first a simplified (closed) system, an ideal quantum gas composed of N noninteracting particles. The basis state can be chosen to be $\vert r_1,r_2,\cdots,r_N\rangle$, but it is more convenient to work in Fock space. Suppose there is a series of energy levels, each one has an energy $\epsilon_i$ and $n_i$ particles on it. Then the state of the system can be expressed in terms of basis state $\vert n_1,n_2,\cdots\rangle\equiv\vert \{n_i\}\rangle$, i.e. $\{\hat{n}_i\}$ is a complete set of observables like Eq.~(\ref{3}) since $[\hat{n}_i,\hat{n}_j]=0$. Besides, we have two conditions in terms of macro-observables
\begin{equation}
\sum_{i}\hat{n}_i=\hat{N},\qquad \sum_{i}\epsilon_i\hat{n}_i=\hat{H},
\label{7}
\end{equation}
i.e. we have a distribution set $\{n_i\}$. Now, turn on some interactions from an environment
\begin{equation}
G_{\hat{U}_{int}}\equiv\{\hat{U}_{int}|[\hat{U}_{int},\hat{N}]\approx0,[\hat{U}_{int},\hat{H}]\approx0\},
\label{8}
\end{equation}
i.e. a collection of \emph{all the possible} interactions $\hat{U}_{int}$ that satisfy the stability conditions as in Eq.~(\ref{5k}). Those interactions are too complicated to be described, but it can be assumed that the average effects can be modelled by an effective interaction $\hat{U}_{\lambda}$ that is a combination of those basic interactions in the set $G_{\hat{U}_{int}}$, i.e. $\hat{U}_{\lambda}\approx\hat{U}_1\hat{U}_2\cdots$ with $\lambda$ a (adiabatic) parameter. Furthermore we assume that, as a result of $\hat{U}_{\lambda}$, we have a new basis state for the system given by
\begin{equation}
\vert \{n_i(\lambda^i_\alpha)\}\rangle, \qquad \lambda^i_\alpha=\lambda^i_1,\cdots,\lambda^i_{g_i},
\label{9}
\end{equation}
i.e. each of the original energy levels is split into some sub-levels $\epsilon_i(\lambda^i_\alpha)$ with corresponding particle number $n_i(\lambda^i_\alpha)$ on it, with $g_i$ the number of sub-levels on each original level $\epsilon_i$. The changes of states can be formally expressed as a change of the system's Hilbert space, $\mathcal{H}_{\{n_i\}}\rightarrow\mathcal{H}_{\{n_i(\lambda)\}}$. The operation of $\hat{U}_{\lambda}$ on a state $\vert\Psi\rangle$ in $\mathcal{H}_{\{n_i\}}$ can be expressed as
\begin{equation}
\hat{U}_{\lambda}\vert\Psi\rangle\vert\chi\rangle_E=\sum_{\{n_i\}}C_{\{n_i\}}\hat{U}_{\lambda}\vert \{n_i\}\rangle\vert\chi\rangle_E=\sum_{\{n_i(\lambda)\}}D_{\{n_i(\lambda)\}}\vert \{n_i(\lambda)\}\rangle\vert\chi(\lambda)\rangle_E,
\label{10}
\end{equation}
i.e. the system is correlated with the environment denoted by the parameter $\lambda$, with $\vert\cdot\rangle_E$ a state for the environment. After partially tracing over the environment~\cite{o,p,p1}, we will obtain a collection of states in which each one has a fixed distribution set $\{n_i(\lambda)\}$, leading to a ``quasi-ensemble". We can further collect the sub-levels $\lambda^i_\alpha,\alpha=1,2,\cdots,g_i$ for each $n_i$ to obtain a \emph{coarse grained} ``quasi-ensemble", with each member's space denoted as $\mathcal{F}\{n_i\}$\footnote{Note that, the particle distribution $\{n_i\}$ here is different from the one in Eq.~(\ref{5g}) which is some probability distribution for the states in the ``quasi-ensemble".}. Then, we have an effective reduction\footnote{The probability for each distribution $\{n_i\}$ can be randomized to be equal approximately, due to those complicated interactions.}
\begin{equation}
\mathcal{H}_{\{n_i(\lambda)\}}\rightarrow\bigoplus_{\{n_i\}}\mathcal{F}\{n_i\},
\label{12}
\end{equation}
with $\bigoplus$ denoted as the collection of the members in the coarse grained ``quasi-ensemble".

The above analysis implies that the space $\mathcal{F}\{n_i\}$ is almost invariant or stable under those operators in the set $G_{\hat{U}_{int}}$.
But those operations always involve the environment. To subtract the environment, we can restrict $G_{\hat{U}_{int}}$ further to a smaller set whose operators only act on the space $\mathcal{F}\{n_i\}$. In the present model, there is a group of particle permutations\footnote{It should be stressed that the permutations must be induced by some actual interactions in nature, although they can be represented by some abstract operators.}among the sub-levels of each energy level when the indistinguishable property of quantum particles is considered, i.e.
\begin{equation}
G_p\{n_i\}\equiv\{\hat{S}_p|[\hat{S}_p,\hat{N}]\approx0,[\hat{S}_p,\hat{H}]\approx0\},
\label{14}
\end{equation}
which also depends on the distribution set $\{n_i\}$.
Therefore, we can define an effective macroscopic space as an approximate quotient space
\begin{equation}
\mathcal{F}_{mac}\{n_i\}\equiv\mathcal{F}\{n_i\}/G_p\{n_i\},
\label{15}
\end{equation}
giving a further \emph{coarse grained} description of the system in terms of macrostates described by approximate equivalence classes. From Eq.~(\ref{15}), we have
\begin{equation}
\dim(\mathcal{F}\{n_i\})/\dim(\mathcal{F}_{mac}\{n_i\})=\dim(G_p\{n_i\}),
\label{16}
\end{equation}
i.e. \emph{$\dim(G_p\{n_i\})$ is the (average) number of microstates per macrostate}. This leads to the Boltzmann entropy for the ``quasi-ensemble" in Eq.~(\ref{12})
\begin{equation}
S_{Bolt}\approx k_B\ln \left[\sum_{\{n_i\}}\dim(G_p\{n_i\})\right],
\label{17}
\end{equation}
which is the entropy (increment) for the system induced by the interactions. This is one part of of the full change $k_B\ln[\dim(G_{\hat{U}_{int}})]$, just like the one in Eq.~(\ref{26e}). The next step is to find the most probable distribution that maximizes the Boltzmann entropy (increment) together with the some restrictions on the energy and particle number, just like Eq.~(\ref{27a}), giving the familiar distribution for the indistinguishable quantum particles\footnote{Note that, although the energy levels are split into sub-levels by the interactions, the macro-observables in Eq.~(\ref{7}) are almost unchanged. Thus, the energy and particle number can still be expressed approximately in terms of the distribution $\{n_i\}$.}.

The above derivation is different from the one in a standard textbook~\cite{r}, where the latter case with (sub-)levels $\{\epsilon_i(\lambda)\}$ is treated as a microcanonical ensemble for a closed system. However, if treated as a (quantum) closed system, it can be well described by quantum mechanics in Fock space representation, with a Hamiltonian $\hat{H}\{\hat{n}_i(\lambda)\}$. In particular, the permutational properties can be well described by the commutative relations of the creators and annihilators $[\hat{a}_i(\lambda^i_\alpha),\hat{a}_j^{\dag}(\lambda^j_\beta)]_{\mp}=\delta_{ij}\delta_{\alpha\beta}$. In this sense, the enumeration of microstates due to the permutations in the textbook's derivation is artificial due to a priori coarse grained description of the (sub-)levels $\{\epsilon_i(\lambda)\}$. In our derivation, the Boltzmann entropy related to the permutations comes from the weak interactions with an environment, with which the system is in equilibrium according to the (stability) conditions in Eq.~(\ref{8}). This gives a grand canonical ensemble for the system open to an environment, similarly for canonical ensemble if the condition for particle number is replaced by $[\hat{U}_{int},\hat{N}]=0$.

\section*{Acknowledgments}
This work is supported by the NNSF of China, Grant No. 11375150.


\end{document}